\documentstyle[rotate,epsf]{mn}

\newcommand{\be}{\begin{equation}}
\newcommand{\ee}{\end{equation}}

\def\Msol{{M_\odot}}

\def\spose#1{\hbox to 0pt{#1\hss}} 
\def\lta{\mathrel{\spose{\lower 3pt\hbox{$\sim$}}\raise 2.0pt\hbox{$<$}}}
\def\gta{\mathrel{\spose{\lower 3pt\hbox{$\sim$}}\raise 2.0pt\hbox{$>$}}}

\begin{document}

\title[Microlensing to M31]{Expectations from Realistic Microlensing
Models of M31. I: Optical Depth}

\author[Geza Gyuk and Arlin Crotts]{Geza Gyuk$^1$ and Arlin Crotts$^2$\\
\\
$^1$Department of Physics, University of California, San Diego, 9500
Gilman Drive, La Jolla, CA 92093\\
$^2$Department of Astronomy, Columbia University, 550 W. 120th St., New York,
NY 10027}

\date{Received ***}
\maketitle

\begin{abstract}
	We provide a set of microlensing optical depth maps for
M31. Optical depths towards Andromeda were calculated on the basis of a
four component model of the lens and source populations: disk and bulge
sources lensed by bulge, M31 halo and Galactic halo lenses. We confirm the
high optical depth and the strong optical depth gradient along the M31
minor axis due to a dark halo of lenses and also discuss the magnitude of
the self-lensing due to the bulge. We explore how the shape of the optical
depth maps to M31 vary with the halo parameters core radius and
flattening.
\end{abstract}

\section{Introduction}

	The ongoing microlensing observations towards the LMC and SMC have
provided extremely puzzling results. On the one hand, analysis of the
first two years of observations (Alcock et al.~1997a) suggest a halo composed
of objects with mass $\sim 0.5\Msol$ and a total mass in MACHOs out to 50
kpc of around $2.0 \times 10^{11}\Msol$.  On the other hand, producing
such a halo requires extreme assumptions about star formation, galaxy
formation, and the cosmic baryonic mass fraction. An attractive
possibility is that the microlenses do not reside in the halo at all!
Alternative suggested locations are the LMC halo (Kerins \& Evans 1999),
the disk of the LMC itself (Sahu 1994), a warped and flaring Galactic disk
(Evans et al.~1998), or an intervening population (Zhao 1998).
Unfortunately, the low event rates, uncertainties in the Galactic model,
and the velocity-mass-distance degeneracy in microlensing all conspire to
make precise determinations of the MACHO parameters difficult. Over the
next decade, second generation microlensing surveys, monitoring ten times
the number of stars in the LMC will improve the overall statistics (and
numbers of ``special'' events) considerably, allowing an unambiguous
determination of the location of the microlenses. Even so, the
paucity of usable lines of sight within our halo makes determination of
the halo parameters such as the flattening or core radius very difficult.

	The Andromeda Galaxy (M31) provides a unique laboratory for
probing the structure of galactic baryonic halos (Crotts 1992). Not only
will the event rate be much higher than for LMC lensing, but it will be
possible to probe a large variety of lines of sight across the disk and
bulge and though the M31 halo.  Furthermore, it provides another example
of a bulge and halo which can be studied, entirely separate from the
Galaxy.  Recently, two collaborations, MEGA and AGAPE, have begun
observations looking for microlensing in the stars of M31. Previous papers
have made it clear that a substantial microlensing signal can be expected.
In this paper we calculate, using realistic mass models, optical depth
maps for M31. The results suggest that we should be able to definitively
say whether M31 has a dark baryonic halo with only a few years or less of
microlensing data. We also discuss how their variation with halo
parameters may allow us to determine the M31 halo structure. This is
particularly important in evaluating the level of resources that should be
dedicated towards the ongoing observational efforts. Preliminary results
suggest that the core radius and density profile power-law should be the
easiest parameters to extract.
	
	The paper is organized in the following manner. In the next
section we briefly discuss the M31 models we used. Following this we
present optical depth maps for various halo models, discuss the
microlensing backgrounds and finish with a quick discussion of the
implications of the maps.

\section{Modeling}

Sources are taken to reside in a luminous two-component model of M31
consisting of an exponential disk and a bulge.  The disk model is inclined
at an angle of 77$^\circ$ and has a scale length of 5.8 kpc and a central
surface brightness of $\mu_R = 20$ (Walterbos \& Kennicutt 1988).  The
bulge model is based on the ``small bulge'' of Kent (1989) with a central
surface brightness of $\mu_R = 14$. This is an axisymmetric bulge with a
roughly exp(-$r^{0.4}$) falloff in volume density with an effective radius
of approximately 1 kpc and axis ratio, $c/a \sim 0.8$. Values of the bulge
density are normalized to make $M_{bulge} = 4 \times 10^{10} \Msol$.

The predominant lens population is taken to be the M31 dark matter halo.
We explore a parametrized set of M31 halo models. Each model halo is a cored
``isothermal sphere'' determined by three parameters: the flattening ($q$), the
core radius ($r_c$) and the MACHO fraction ($f_b$):
\be
\rho(x,y,z) = \frac{V_c(\infty)^2}{4 \pi G}\frac{e}{a^2q \sin^{-1} e}
\frac{1}{x^2+y^2 + (z/q)^2 +a^2},
\ee
where a is the core radius, q is the x-z axis ratio, $e=\sqrt{1-q^2}$ and
$V_c(\infty)=240$km/s is taken from observations of the M31 disk. In
section 4 we briefly consider the optical depth due to other populations
such as the bulge stars.

More details of our modeling are given in Gyuk \& Crotts (1999) where in
particular the velocity distributions (necessary for calculation of the
microlensing {\em rate}) are discussed.
These considerations do not affect the optical depths treated here.

\section{Optical Depth Maps}

\begin{figure}
\epsfysize=10.0cm
\centerline{
\rotate[r]{\epsfbox{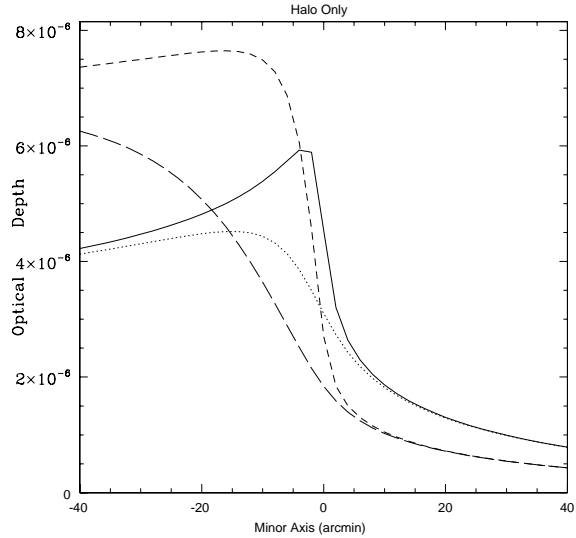}}}
\caption{Halo optical depth along the minor axis. The curves are: solid line --
q=0.3 core=1kpc, dotted line -- q=0.3 core= 5kpc, dashed line -- q=1.0
core=1.0kpc, and long dashed line -- q=1.0 core=5.0kpc }
\label{tauminorslice}
\end{figure}

The classical microlensing optical depth is defined as the number of lenses
within one Einstein radius of the source-observer line-of-sight (the
microlensing tube):
\be
\tau = \int_0^D  \frac{\rho_{\rm halo}(d)}{M_{\rm lens}}\frac{4 G M_{\rm lens}}{c^2}\frac{(D-d)d}{D}
\ee
Such a configuration is intended to correspond to a
``detectable magnification'' of at least a factor of $1.34$. Unfortunately, in
the case of non-resolved stars (``pixel lensing'') we have typically
\be
\pi \sigma^2 S_{M31} >> L_{*}.
\ee 
where $S_{M31}$ is the background surface brightness, $4\pi\sigma^2$ is
the effective area of the seeing disk and $L_{*}$ is the luminosity of
the source star. Thus it is by no means certain that a modest increase of
$L_{*} \rightarrow 1.34 L_{*}$, as the lens passes within an Einstein
radius, will be detectable. Furthermore, even for the events detected,
measurement of the Einstein timescale $t_0$ is difficult. Thus measurement
of the optical depth may be difficult.  Nonetheless, advances have been
made in constructing estimators of optical depth within highly crowded
star fields (Gondolo 1999), which do not require the Einstein timescale
for individual events, although they still require evaluation of the
efficiency of the survey in question for events with various half maximum
timescales.  The errors on the derived optical depths will likely be
larger than for the equivalent number of classical microlensing events. It
is clear, however, that image image subtraction techniques (Tomaney \&
Crotts 1996, Alcock et al.~1999a, b) can produce a higher event rate than
conventional photometric monitoring.  Thus one needs models of the
optical depth, even if expressed only in terms of the cross-section for a
factor 1.34 amplification in order to understand how microlensing across
M31 will differ depending on the spatial distribution of microlensing
masses in the halo and other populations.

\begin{figure}
\epsfysize=10.0cm
\centerline{
\rotate[r]{\epsfbox{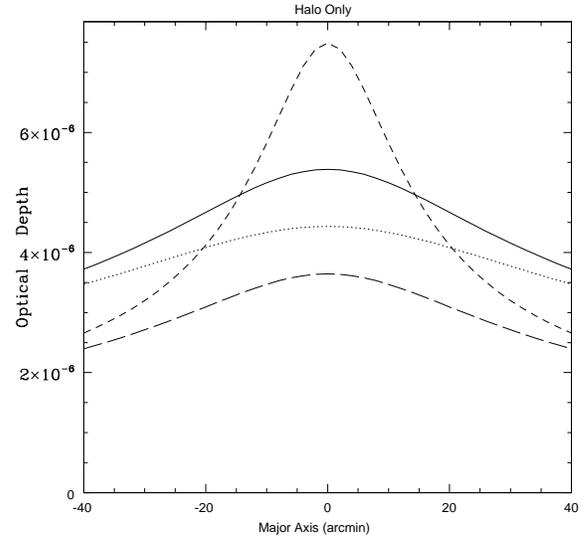}}}
\caption{Optical depth along a line parallel to the major axis and offset along
the minor axis by -10$^\prime$ (towards the far side of the disk).
The curves are: solid
line -- q=0.3 core=1kpc, dotted line -- q=0.3 core= 5kpc, dashed line --
q=1.0 core=1.0kpc, and long dashed line -- q=1.0 core=5.0kpc }
\label{taumajorslice}
\end{figure}

The above expression for the optical depth must be slightly amended to
include the effects of the three-dimensional distribution of the source
stars, especially of the bulge. We thus integrate the source density along
the line of sight giving
\be
\tau = \frac{\int_0^\infty \rho(S) \int_0^S  \frac{\rho_{\rm halo}(s)}{M_{\rm
lens}}\frac{4 G M_{\rm lens}}{c^2}\frac{(S-s)s}{S} ds dS}
{\int_0^\infty \rho(S) dS}
\ee
\begin{figure*}
\centerline{
\hbox{
\hfill
\vbox{
\epsfysize=10.0cm
\rotate[r]{\epsfbox{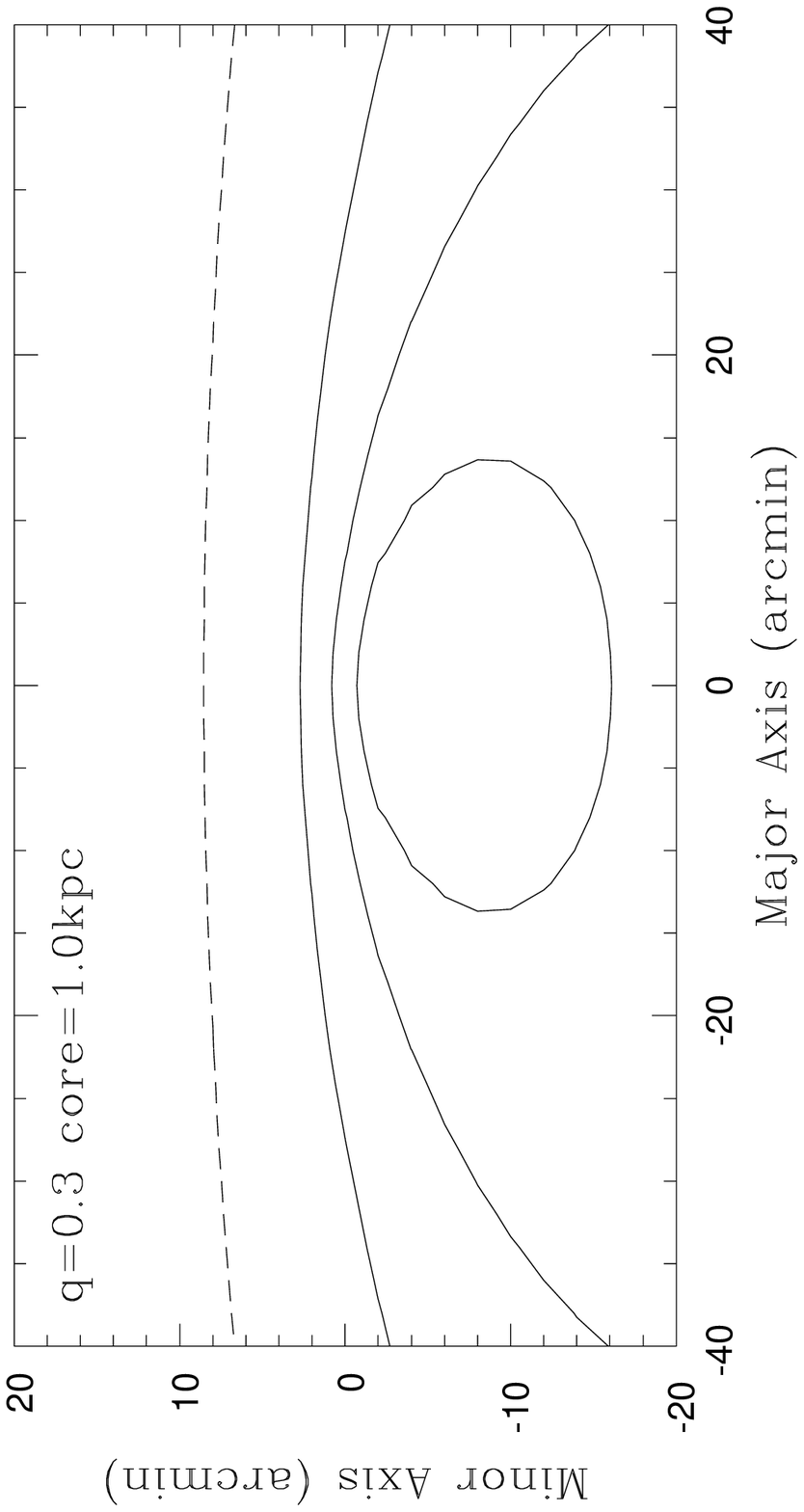}}
\epsfysize=10.0cm
\rotate[r]{\epsfbox{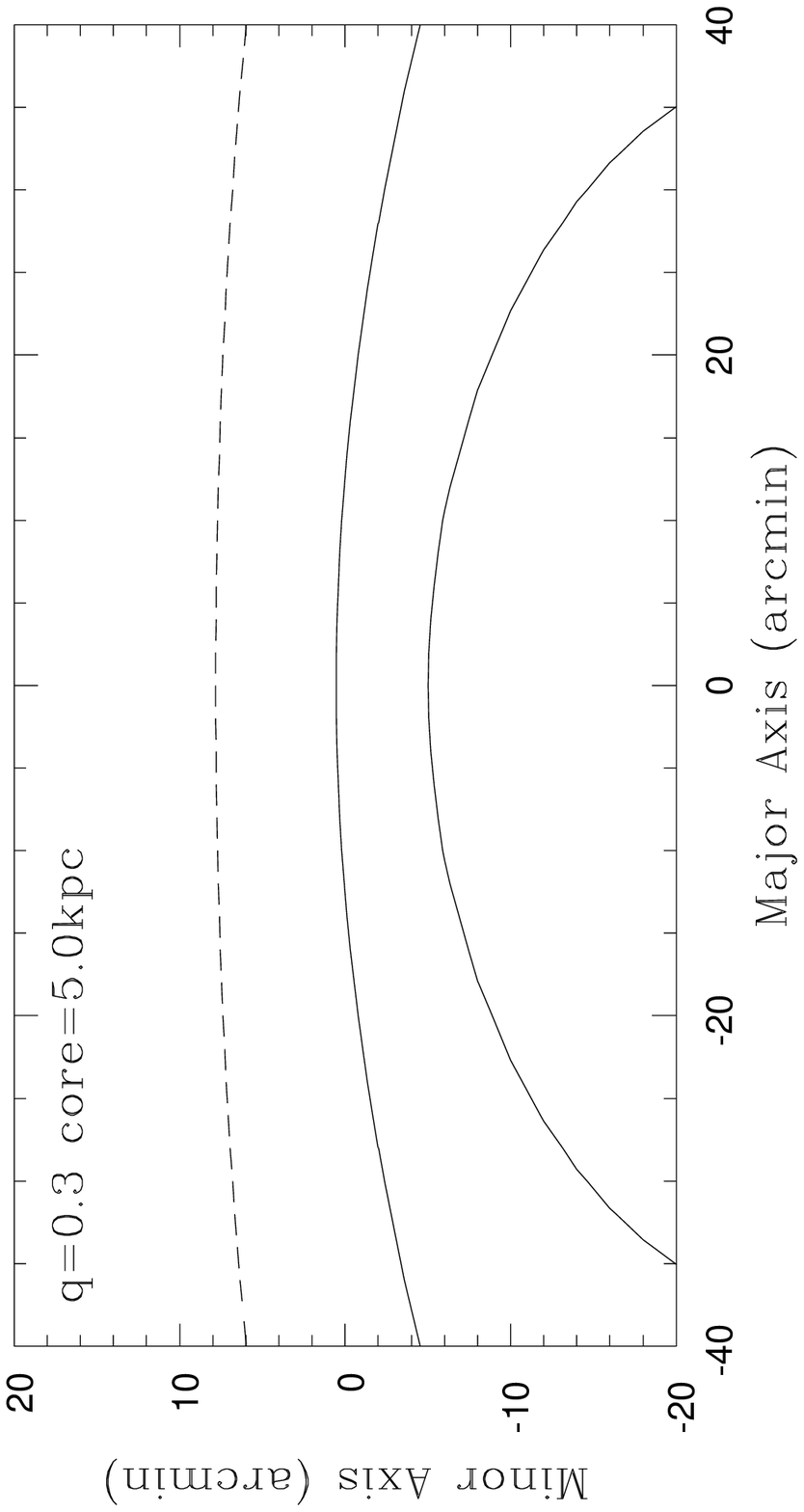}}}
\vbox{
\epsfysize=10.0cm
\rotate[r]{\epsfbox{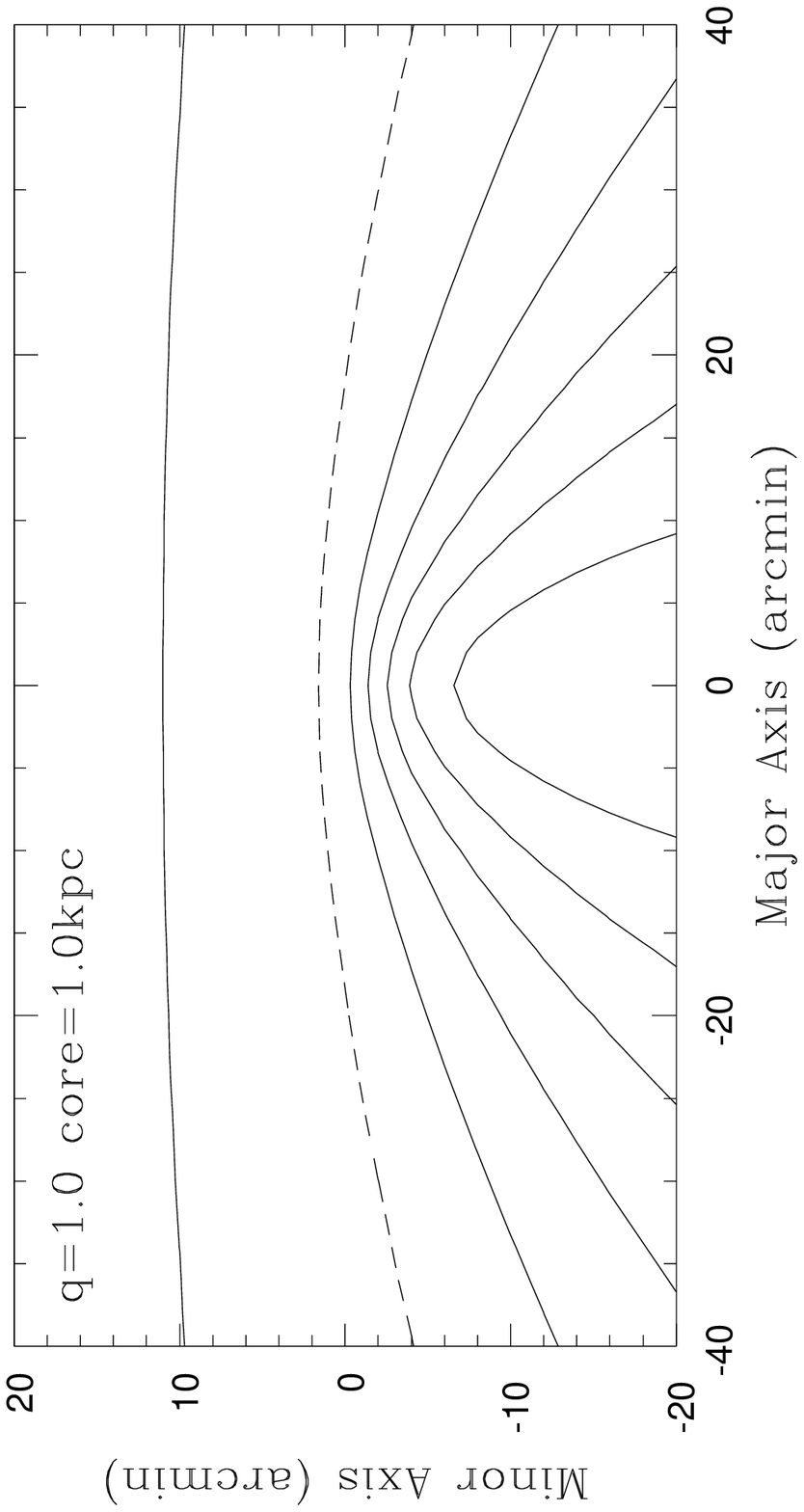}}
\epsfysize=10.0cm
\rotate[r]{\epsfbox{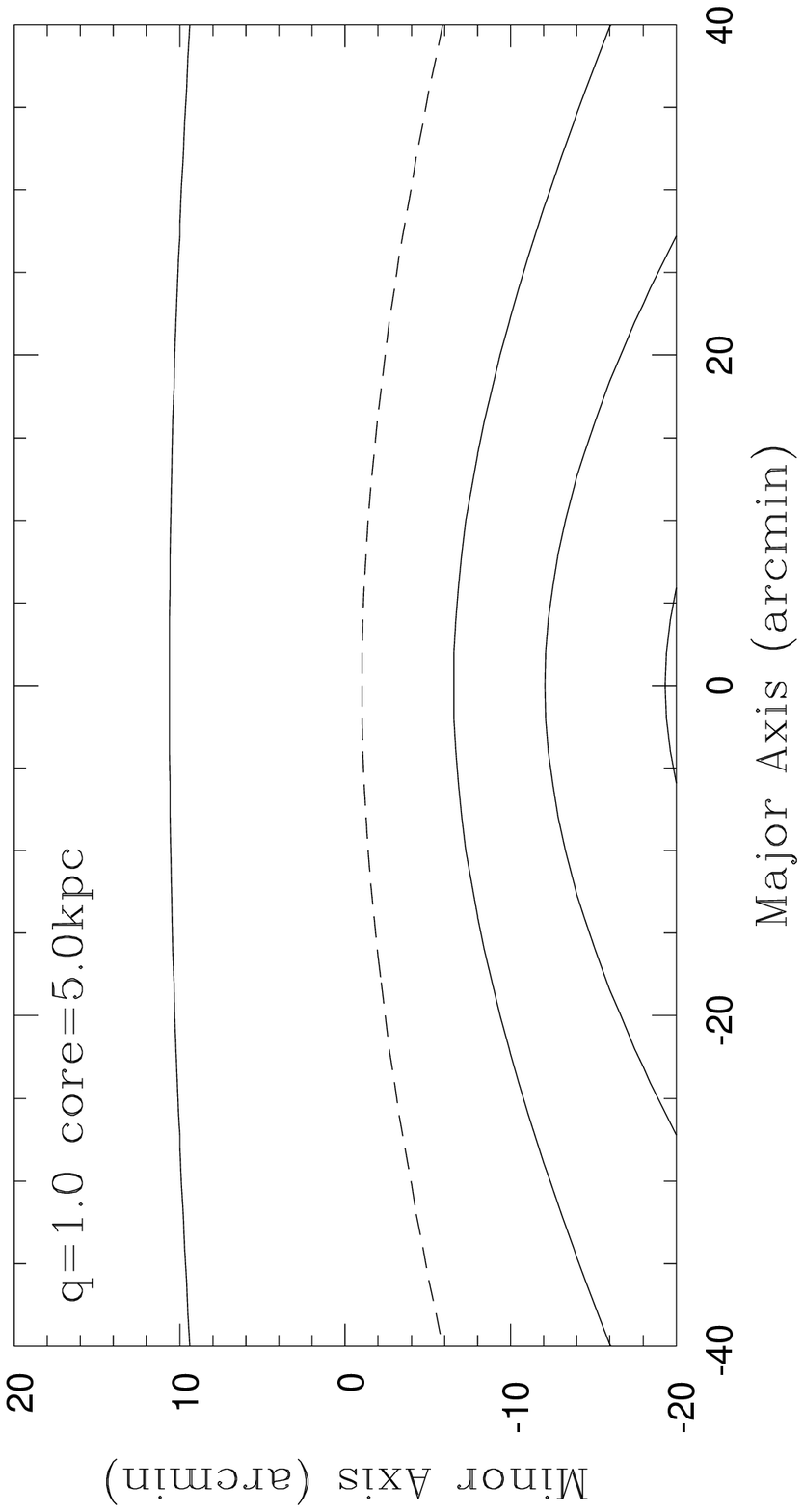}}}
\hfill
}}
\caption{Contours of optical depth for halo models a) q=0.3 core=1.0,
b) q=0.3 core=5.0, c) q=1.0 core=1.0 halo and d)q=1.0 core=5.0. Contours
are from top to bottom: a) 2,3,4 and 5 $\times 10^{-6}$, b) 2,3 and 4
$\times 10^{-6}$, c) 1,2,3,4,5,6 and 7 $\times 10^{-6}$ and d) 1,2,3,4 and
5 $\times 10^{-6}$. For all models the dashed contour is 2.0$\times 10^{-6}$. }
\label{taumodels}
\end{figure*}
The results of this calculation as a function of position for a variety of
halo models are shown in Figure \ref{taumodels}.  The most important
attribute is the strong modulation of the optical depth from the near to
far side of the M31 disk as was first remarked on by Crotts
(1992). Near-side lines-of-sight have considerably less halo to penetrate
and hence a lower optical depth. This can be seen nicely in Figure
\ref{tauminorslice} where we plot the optical depth along the minor axis
for the four models depicted in Figure \ref{taumodels}. While all models
exhibit the strong variation from near to far, the fractional variation in
$\tau$ across the minor axis is most pronounced for less flattened models, and
changes in $\tau$ along the minor axis occur most rapidly for models with small
core radii.
This can be understood geometrically: in the limit of an
extremely flattened halo the pathlength (and density run) through the halo
is identical for locations equidistant from the center.  Small core radii
tend to make the central gradient steeper and produce a maximum at a
distance along the minor axis comparable to the core size. This maximum is
especially prominent in the flattened halos.

Variations in core radii and flattening are also reflected in the run of
optical depth along the major axis. In Figure \ref{taumajorslice} we show
the optical depth along the major axis displaced by -10$^\prime$ on the minor
axis. The gradients in the small core radii models are much larger than
for large core radii. Asymptotically the flattened halos have a larger
optical depth.

\section{Background Lensing}

\begin{figure}
\epsfysize=10.0cm
\centerline{
\rotate[r]{\epsfbox{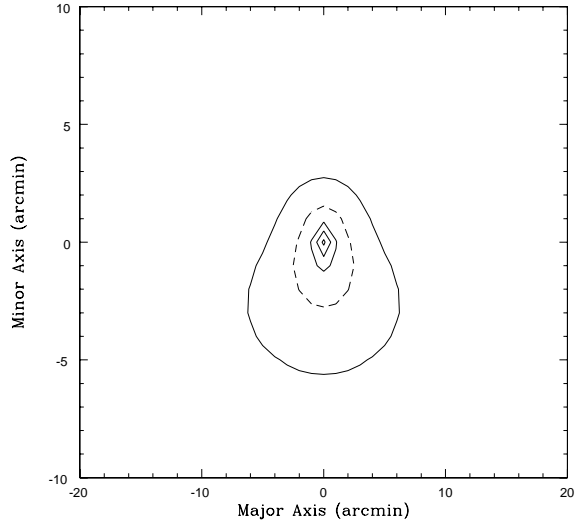}}}
\caption{Contours of optical depth for the bulge self-lensing. Contours
are, from the outside in: 1,2,3,4 and 5 $\times 10^{-6}$. Note that the
region shown is half the dimensions of the maps of Figure 3.}
\label{bulgelens}
\end{figure}

Unfortunately, the M31 halo is not the only source of lenses. As mentioned
above, the bulge stars can also serve as lenses. We show in Figure
\ref{bulgelens} the optical depth contributed by the bulge lenses. The
effect of the bulge lenses is highly concentrated towards the center. This
is a mixed blessing. On the one hand the bulge contribution can thus be
effectively removed by deleting the central few arcminutes of M31. Beyond
a radius of 5 arcminutes, bulge lenses contribute negligibly to the overall
optical depth. On the other hand the source densities are much higher in
the central regions and thus we expect the bulk of our halo events to
occur in these regions. We discuss this point in more detail in a
forthcoming paper (Gyuk \& Crotts 1999).
The bulge of M31 might easily serve as an interesting foil to the Galactic
Bulge, which produces microlensing results which seem to require a special
geometry relative to the observer, or other unexpected effects (Alcock et
al.~1997b, Gould 1997).

In addition to the M31 bulge lensing a uniform optical depth across the
field will be contributed by the Galactic halo. This contribution will be
of order $\sim 10^{-6}$ corresponding to a 40\% Galactic halo as suggested
by the recent LMC microlensing results. Finally, disk self lensing will
occur. The magnitude of the optical depth for this component will however
be at least an order of magnitude lower than the expected halo or bulge
contributions (Gould 1994) and hence is ignored in these calculations.

\section{Discussion and Conclusions}

The optical depth maps for M31 shown above exhibit a wealth of structure
and clearly contain important information on the shape of the M31
halo. The most important of these information bearing characteristics is
the asymmetry in the optical depth to the near and far sides of the M31
disk. A detection of strong variation in the optical depth from front to
back will be a clear and unambiguous signal of M31's microlensing halo, perhaps
due to baryons. No
other lens population or contaminating background can produce this
signal. However, the lack of a strong gradient should not be taken as
conclusive proof that M31 does not have a halo. As discussed above, strong
flattening or a large core radius can reduce or mask the
gradient. Nevertheless, the halo should still be clearly indicated by the
high microlensing rates observed outside the bulge region. In such a case,
however, careful modeling of the experimental efficiency and control over
the variable star contamination will be necessary to insure that the
observed event are really microlensing.

Further information about the structure of the M31 baryonic halo can 
be gleaned from the distribution of microlensing along the major
axis. A strong maximum at the minor axis is expected for small core radii
especially for spherical halos.

The combination of the change in event rate both along the major and minor axis
directions can in principle reveal both the core radius and flattening from a
microlensing survey.
How easily such parameters can be measured depends critically on the rate at
which events can be detected, which we discuss in paper II of this series,
along with estimates of the expected accuracy.
Additionally, we will discuss strategies to optimize such surveys for
measuring shape parameters.

\end{document}